\documentclass[prl,aps,twocolumn,showpacs,preprintnumbers,amsmath,amssymb,superscriptaddress]{revtex4-2}
\usepackage{dcolumn}
\usepackage{graphicx}
\usepackage[colorlinks=true,linkcolor=blue,citecolor=blue]{hyperref}
\usepackage{color}

\begin{document}

\title{Photoemission and Dynamical Mean Field Theory Study of Electronic Correlation in a $t_{2g}^{5}$ Metal of SrRhO$_{3}$ Thin Film}

\author{Yujun Zhang}
\email{zhangyujun@issp.u-tokyo.ac.jp}
\affiliation{Graduate School of Materials Science, University of Hyogo, 3-2-1 Kouto, Kamigori-cho, Ako-gun, Hyogo 678-1297, Japan}
\affiliation{Institute for Solid State Physics, University of Tokyo, 5-1-5 Kashiwanoha, Chiba 277-8581, Japan}
\author{Minjae Kim}
\email{garix.minjae.kim@gmail.com}
\affiliation{Department of Physics and Astronomy, Rutgers University, Piscataway, New Jersey 08854, USA}
\affiliation{Coll\`ege de France, 11 place Marcelin Berthelot, 75005 Paris, France}
\author{Jernej Mravlje}
\affiliation{Jo\v{z}ef Stefan Institute, Jamova 39, Ljubljana, Slovenia}
\author{Changhee Sohn}
\affiliation{Materials Science and Technology Division, Oak Ridge National Laboratory, Oak Ridge, Tennessee 37831, USA}
\author{Yongseong Choi}
\affiliation{Advanced Photon Source, Argonne National Laboratory, Argonne, Illinois 60439, USA}
\author{Joerg Strempfer}
\affiliation{Advanced Photon Source, Argonne National Laboratory, Argonne, Illinois 60439, USA}
\author{Yasushi Hotta}
\affiliation{Department of Engineering, University of Hyogo, 2167 Shosha, Himeji, Hyogo 671-2280, Japan}
\author{Akira Yasui}
\affiliation{Japan Synchrotron Radiation Research Institute (JASRI), 1-1-1 Kouto, Sayo, Hyogo 679-5198, Japan}
\author{John Nichols}
\affiliation{Department of Physics and Astronomy, University of Arkansas at Little Rock, Little Rock, AR, 72204, USA}
\author{Ho Nyung Lee}
\affiliation{Materials Science and Technology Division, Oak Ridge National Laboratory, Oak Ridge, Tennessee 37831, USA}
\author{Hiroki Wadati}
\affiliation{Graduate School of Materials Science, University of Hyogo, 3-2-1 Kouto, Kamigori-cho, Ako-gun, Hyogo 678-1297, Japan}
\affiliation{Institute for Solid State Physics, University of Tokyo, 5-1-5 Kashiwanoha, Chiba 277-8581, Japan}

%\date{\today}
%
\begin{abstract}
Perovskite rhodates are characterized by intermediate strengths of
both electronic correlation as well as spin-orbit coupling (SOC) and
usually behave as moderately correlated metals. A recent publication
(Phys. Rev. B 95, 245121(2017)) on epitaxial SrRhO$_3$
thin films unexpectedly reported a bad-metallic behavior and suggested
the occurrence of antiferromagnetism below 100~K. We studied this 
SrRhO$_3$ thin film by hard \mbox{x-ray} photoemission spectroscopy and found
a very small density of states (DOS) at Fermi level, which is consistent with the reported
bad-metallic behavior. However, this negligible DOS persists up to room
temperature, which contradicts with the explanation of antiferromagnetic transition at around 100 K. We also employed electronic structure calculations within the framework of
density functional theory and dynamical mean-field
theory. In contrast to the experimental results, our
calculations indicate metallic behavior of both bulk SrRhO$_3$ and  the SrRhO$_3$ thin film. The thin film exhibits stronger correlation effects than the bulk, but the correlation effects are not
sufficient to drive a transition to an insulating state. The
calculated uniform magnetic susceptibility is substantially larger in the
thin film than that in the bulk. The role of SOC was also investigated and only a
moderate modulation of the electronic structure was observed. Hence
SOC is not expected to play an important role for electronic
correlation in SrRhO$_3$. 
\end{abstract}
\maketitle
\section{Introduction}

4$d$ transition metal compounds are characterized by the moderate strengths of both
electronic correlation and spin-orbit coupling (SOC) compared to their
strongly correlated 3$d$ and strongly spin-orbit-coupled 5$d$
counterparts. Nevertheless,  4$d$ systems do exhibit interesting physical properties as well. Notable examples are found especially in the 
perovskite ruthenate family: unconventional superconductivity in
Sr$_2$RuO$_4$~\cite{1_maeno1994superconductivity,2_mackenzie2003superconductivity},
ferromagnetism (FM) in SrRuO$_3$~\cite{3_kanbayasi1976magnetic}, and
current-induced insulator-metal transition in Ca$_2$RuO$_4$~\cite{4_nakamura2013electric,5_sow2017current}, etc. Magnetism plays an important role in many 4$d$ systems. 
 Spin-triplet superconductivity in Sr$_2$RuO$_4$
was argued to be related to ferromagnetic spin fluctuations~\cite{1_maeno1994superconductivity,2_mackenzie2003superconductivity}. FM
is realized in SrRuO$_3$~\cite{3_kanbayasi1976magnetic} while
CaRuO$_3$~\cite{6_cao1997thermal} and Sr$_3$Ru$_2$O$_7$~\cite{7_cao1997observation,8_ikeda2000ground} are presumably close to FM. On
the other hand, Ca$_2$RuO$_4$ is a special case that exhibits an
antiferromagnetic (AFM) insulating ground state and insulator-metal
transition~\cite{9_nakatsuji1997ca2ruo4,10_carlo2012new}. However, generally magnetic ordering is rarely observed in other perovskite 4$d$ oxides~\cite{11_oka2015intrinsic,12_wadati2014photoemission}. 

Since Rh is the neighbour of Ru in the 4$d$ transition metal series,
Rh-based perovskite oxides, such as SrRhO$_3$~\cite{13_yamaura2001enhanced,14_singh2003prospects,15_yamaura2003electronic,16_nichols2017electronic},
Sr$_2$RhO$_4$~\cite{17_perry2006sr2rho4,18_haverkort2008strong,19_liu2008coulomb,20_martins2011reduced}
and Sr$_3$Rh$_2$O$_7$~\cite{21_yamaura2002crystal}, have also
attracted considerable research attention. In the bulk state, these
rhodates are usually correlated metals without magnetic
ordering. Among them, SrRhO$_3$ has the most simple
crystal structure. As first reported by Yamaura \textit{et al.}~\cite{13_yamaura2001enhanced}, bulk SrRhO$_3$ has a GdFeO$_3$-type
distorted perovskite structure with space group \textit{Pnma}. Metallic transport behavior was observed down to 1.8~K~\cite{13_yamaura2001enhanced} and covalent doping of Ca at the Sr-site
does not have significant influence on the metallic state of SrRhO$_3$~\cite{14_singh2003prospects}. Nevertheless, there are several reports
that strongly indicate the instability towards magnetic ordering in
SrRhO$_3$. An enhanced paramagnetic susceptibility~\cite{13_yamaura2001enhanced}
and related theoretical investigations~\cite{15_yamaura2003electronic}
indicate that SrRhO$_3$ is near a quantum critical point with
significant ferromagnetic quantum fluctuation. 

Recently, epitaxial
SrRhO$_3$ thin films were successfully synthesized and their transport and
magnetic properties were reported by Nichols \textit{et al.}~\cite{16_nichols2017electronic}. No FM was observed in the SrRhO$_3$ thin
films but subtle anomalies appeared at around 100~K in magnetization
and magnetoresistance, which indicates the possibility of a magnetic
transition. Based on density functional theory (DFT) +$U$ calculations, Ref.~\cite{16_nichols2017electronic}
suggested that a C-type AFM structure is energetically favorable.  One remarkable result from Ref.~\cite{16_nichols2017electronic} is 
 that the resistivity of SrRhO$_3$ is
very weakly dependent on temperature at high temperature and exhibits an upturn upon cooling below 100 K, showing a weakly
insulating behavior.  Although this behavior is probably related to some AFM order, there is no direct experimental
confirmation about this point so far.

In the present work, we studied SrRhO$_3$ epitaxial thin film by hard x-ray
photoemission spectra (HAXPES) to characterize its electronic structure and the correlation effects, as was earlier done for other perovskite transition metal oxides in Refs.~\cite{22_takizawa2005manifestation,23_sekiyama2004mutual,24_takizawa2009coherent,12_wadati2014photoemission}.
Instead of a coherent peak, a negligible density of states (DOS) near Fermi
level ($E_F$) was observed. This is the case for the results both at room temperature and at 80~K, which precludes the
interpretation of the resistivity upturn upon cooling for temperature
around 100~K in terms of the gap opening induced by AFM
order. Realistic dynamic mean field theory (DMFT) calculations were
conducted to investigate the electronic correlation and instability
towards magnetic ordering in the SrRhO$_3$ thin film. Since  SOC can also play a significant role to influence the
electronic structure of 4$d$ perovskite oxides such as Sr$_2$RuO$_4$~\cite{18_haverkort2008strong,25_kim2018spin} and Sr$_2$RhO$_4$~\cite{18_haverkort2008strong,19_liu2008coulomb,20_martins2011reduced},
the possible effects of SOC are investigated and discussed as well.

\section{Methods}

A 9 nm-thick epitaxial SrRhO$_3$ thin film was grown on a SrTiO$_3$(001) single crystalline substrate by pulsed laser deposition. The details of the fabrication methods and basic characterization of the sample were previously reported in Ref.~\cite{16_nichols2017electronic}.

HAXPES of the SrRhO$_3$ thin film was measured at BL47XU of SPring-8. The incidence angle of linearly polarized ($\pi$-polarization) 7.94~keV hard \mbox{x-ray} was set at 1\textsuperscript{o} and photoemission spectra were collected by a Scienta \mbox{R-4000} electron energy analyzer with an energy resolution of around 250~meV. Surface-sensitive soft \mbox{x-ray} photoemission spectra (SXPES) were measured by a PHI 5000 VersaProbe system (ULVAC-PHI Inc.) with perpendicular incidence of Al $K\alpha$ radiation (1468.7~eV). The energy resolution of the SXPES measurement was around 450~meV. The position of $E_F$ and the energy resolution of both photoemission measurements were determined by measuring and fitting the spectra of a Au reference sample, which was in electrical contact with the SrRhO$_3$ thin film. For temperature dependent HAXPES measurement, a He-flow cryostat was employed to cool the sample down to 80 K.

X-ray linear dichroism (XLD) and resonant magnetic \mbox{x-ray} diffraction (RMXD) measurements of the SrRhO$_3$ thin film at the Rh $L$ edge were carried out at beamline \mbox{4-ID-D} of Advanced Photon Source. For the room temperature XLD measurement, linearly polarized \mbox{x-rays} with electric field component \textit{E} perpendicular and parallel to the sample surface were utilized to measure the \mbox{x-ray} absorption spectra (XAS). The incidence angle of the \mbox{x-rays} was set at around 3\textsuperscript{o} and the XAS signal was collected by partial fluorescence yield mode. RXMD measurement was conducted at 30~K by cooling the sample with an ARS He Displex cryocooler.

For the DFT calculation, we used the Wien2k package~\cite{26_blaha2001wien2k} and local density approximation (LDA) was employed for the calculation of the exchange correlation potential.
For the DFT+DMFT calculation, we used the TRIQS framework~\cite{27_aichhorn2016triqs,28_parcollet2015triqs,29_seth2016triqs,30_gull2011continuous} and treated $t_{2g}$ orbitals by using a rotationally invariant Kanamori Hamiltonian with parameters $U=2.3$~eV and $J=0.4$~eV, which has been used to precisely describe the neighbouring ruthenates~\cite{31_mravlje2011coherence,32_dang2015electronic}.
To compute the magnetic susceptibility by DFT+DMFT, we applied a magnetic field of 5~meV (around 86~T) and allowed symmetry breaking
of spin up and down. For crystal structure optimization of the SrRhO$_3$ thin film, we assumed the in-plane symmetry and lattice constants of the SrTiO$_3$ substrate and relaxed the internal positions of atoms in the unit cell by using DFT~\cite{16_nichols2017electronic}.

\section{Photoemission Results}

%%%%%%%%%%%%%%%%%%%%%%%%%%%%%%%%%%%%%%%%%%%%%%%%%%%%%%%%%%%%%%%%
\begin{figure}[t]
	\includegraphics[width=\columnwidth]{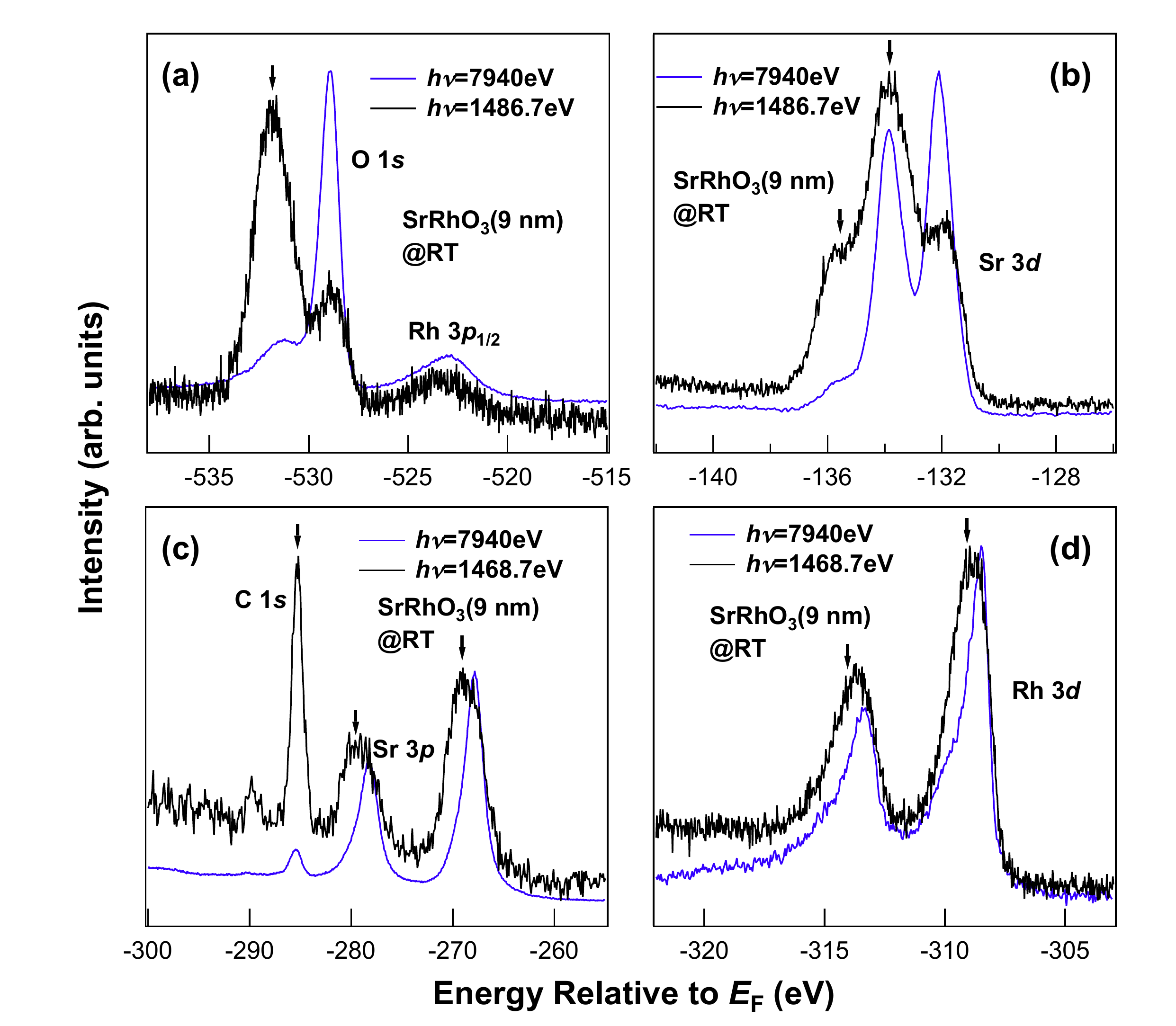}
	\caption{Room temperature (a) O 1$s$ and Rh 3$p_{1/2}$; (b) Sr 3$d$; (c) C 1$s$ and Sr 3$p$; (d) Rh 3$d$ core level photoemission spectra of the SrRhO$_3$ thin film. The arrows indicate the surface components, which are enhanced in SXPES.
		\label{fig:Core level}
	}
\end{figure}
%%%%%%%%%%%%%%%%%%%%%%%%%%%%%%%%%%%%%%%%%%%%%%%%%%%%%%%%%%%%%%%%
Fig.\ref{fig:Core level} shows the core level HAXPES and SXPES results of the SrRhO$_3$ thin film. HAXPES is quite bulk-sensitive and the detection depth is beyond the thickness of the film since the photoemission signal of Ti in the substrate can be observed. On the other hand, SXPES is very surface sensitive, whose typical detection depth is around 1 to 2~nm~\cite{33_seah1979quantitative}. It can be noticed in Fig.\ref{fig:Core level} that surface components (-532~eV O 1$s$ peak in Fig.\ref{fig:Core level}(a); -135.7~eV/-133.9~eV Sr 3$d$ peaks in Fig.\ref{fig:Core level}(b); C 1$s$ contamination signal at -285.5~eV in Fig.\ref{fig:Core level}(c); left shoulders of Sr 3$p$ in Fig.\ref{fig:Core level}(c) and left shoulders of Rh 3$d$ in Fig.\ref{fig:Core level}(d)) have different binding energies compared to the bulk components (-528.9~eV O 1$s$ peak in Fig.\ref{fig:Core level}(a); -133.9~eV/-132.1~eV Sr 3$d$ peaks in Fig.\ref{fig:Core level}(b); main peaks of Sr 3$p$ at -278.2~eV/-267.9~eV in Fig.\ref{fig:Core level}(c) and main peaks of Rh 3$d$ at -313.3~eV/-308.5~eV in Fig.\ref{fig:Core level}(d)). The surface components have little influence on the HAXPES results. The binding energies of peaks in the Rh 3$d$ spectra are also consistent with previous reports of Rh\textsuperscript{4+} oxides~\cite{34_le2011electronic}.

%%%%%%%%%%%%%%%%%%%%%%%%%%%%%%%%%%%%%%%%%%%%%%%%%%%%%%%%%%%%%%%%
\begin{figure}[t]
	\includegraphics[width=\columnwidth]{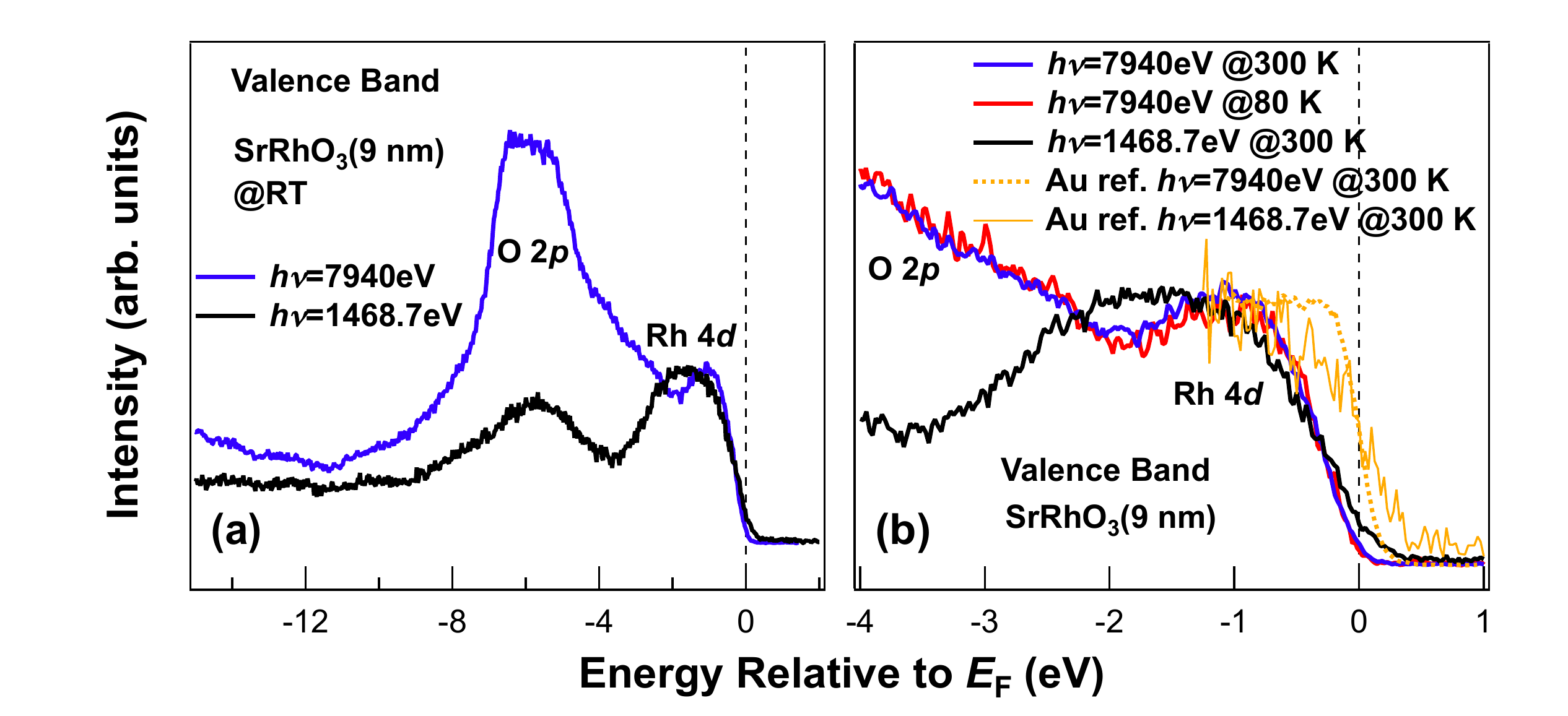}
	\caption{(a) Room temperature valence band photoemission spectra of the SrRhO$_3$ thin film. (b) Enlarged valence band photoemission spectra of the SrRhO$_3$ thin film. The data of reference Au are also shown. 
		\label{fig:Valence band}
	}
\end{figure}
%%%%%%%%%%%%%%%%%%%%%%%%%%%%%%%%%%%%%%%%%%%%%%%%%%%%%%%%%%%%%%%%
The valence band HAXPES and SXPES results are displayed in
Fig.\ref{fig:Valence band}(a). Due to the different photoionization
cross section of $p$ and $d$ levels for hard and soft x-rays~\cite{35_trzhaskovskaya2001photoelectron}, HAXPES is more sensitive to
$p$ levels and SXPES is more sensitive to $d$ levels. By comparing the
HAXPES and SXPES results, it can be concluded that the features in
the energy range from $-10$~eV to $-3$~eV are dominated by O 2$p$ emission
and the features above $-3$~eV mainly come from Rh 4$d$
emission. Fig.\ref{fig:Valence band}(b) shows the valence band spectra
in an expanded region near $E_F$. Surprisingly, the coherent peak is
totally absent for both HAXPES and SXPES. The difference between HAXPES and
SXPES at $E_F$ is mainly due to the different energy resolution of HAXPES and SXPES. By comparing with the corresponding spectra of
the Au reference sample, it is clear that both HAXPES and SXPES have
negligible intensity at $E_F$. It should be noted that the SrTiO$_3$
substrate could also contribute to the HAXPES valence band spectra due
to the large detection depth of HAXPES. However, since SrTiO$_3$ is an
insulator with $d^0$ configuration, it has nearly no contribution to the intensity
above $-3$~eV~\cite{36_haruyama1996angle}.

In Ref.~\cite{16_nichols2017electronic}, the possibility of magnetic
ordering in the SrRhO$_3$ thin film with a transition temperature of
around 100~K was proposed. To investigate the temperature dependence
of the electronic structure in the SrRhO$_3$ thin film, we also conducted
HAXPES measurement at 80 K. However, nearly no temperature dependence
was observed, as shown in Fig.\ref{fig:Valence band}(b). Since
the SrTiO$_3$ substrate has a structural phase transition near this
temperature~\cite{37_ohama1984temperature}, the reported anomalies in
transport properties~\cite{16_nichols2017electronic} may be related to
the change of substrate strain rather than a real magnetic
transition. Consequently, the valence band structure of the SrRhO$_3$ thin film
does not show a significant temperature dependence.

\section{DFT and DFT+DMFT Results}
%%%%%%%%%%%%%%%%%%%%%%%%%%%%%%%%%%%%%%%%%%%%%%%%%%%%%%%%%%%%%%%%
\begin{figure}[t]
	\includegraphics[width=\columnwidth]{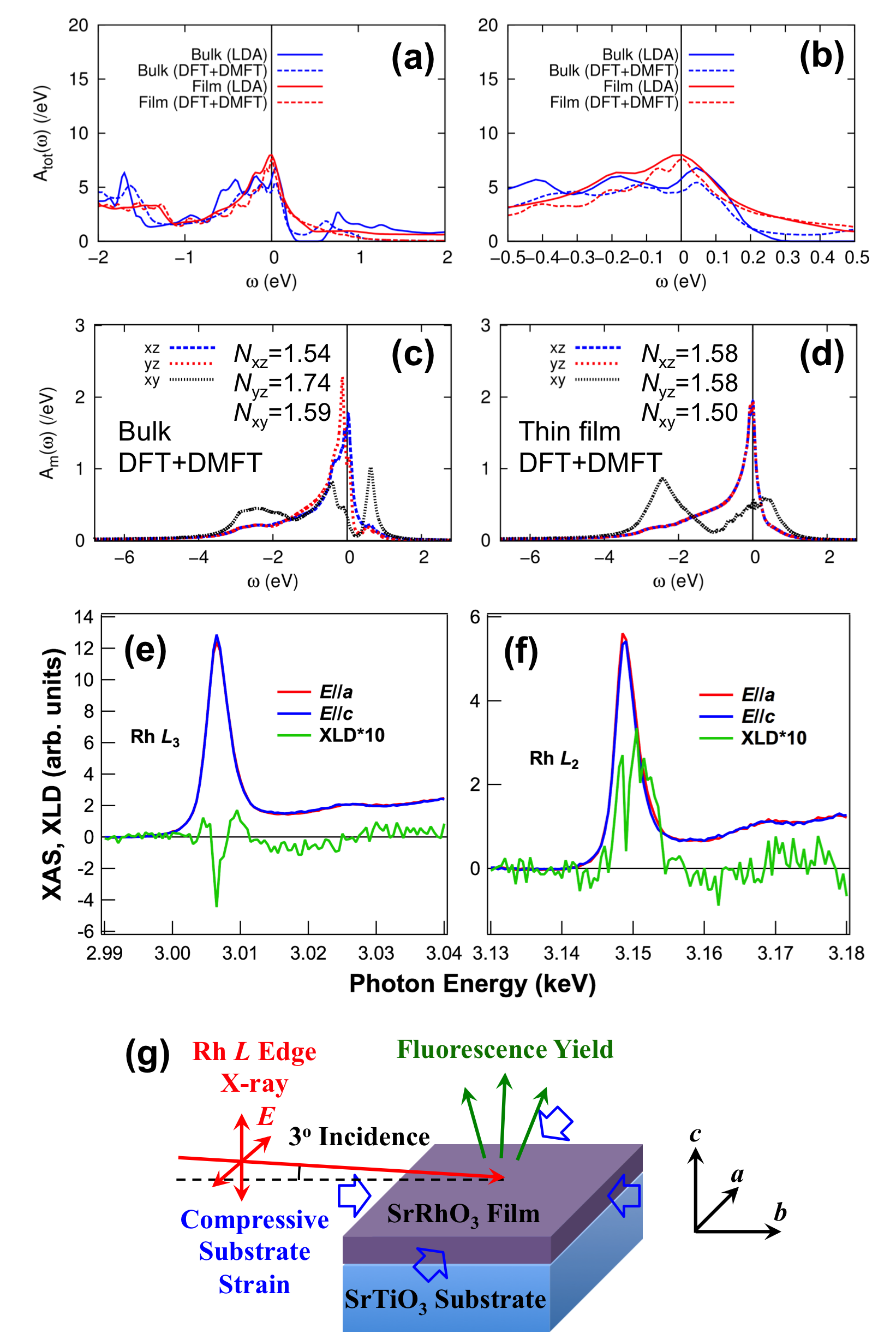}
	\caption{Total DOS $A$\textsubscript{tot}($\omega$) calculated by DFT and DFT+DMFT ($T$=58 K)
		for the bulk and the thin film in (a) wider energy range and (b) narrower energy range. Orbitally resolved DFT+DMFT DOS $A$\textsubscript{m}($\omega$) at $T$=58~K for (c) the bulk and (d) the thin film. $N$ in the figures indicates the orbital occupation. $\omega$ is the energy relative to $E_F$. (e,f) Rh $L$ edge XAS and XLD spectra of the SrRhO$_3$ thin film. (g) Schematic of the XLD measurement for the SrRhO$_3$ thin film under compressive substrate strain.
		\label{fig:Density_of_states}
	}
\end{figure}
%%%%%%%%%%%%%%%%%%%%%%%%%%%%%%%%%%%%%%%%%%%%%%%%%%%%%%%%%%%%%%%%

The photoemission results above suggest negligibly small DOS near $E_F$ in the SrRhO$_3$ thin film, in contrast to bulk SrRhO$_3$ that exhibits metallic behavior~\cite{13_yamaura2001enhanced}. In order to understand this phenomenon, we now turn to the DFT(+DMFT) calculations, performed for both the bulk and the thin film.

\begin{figure}[t]
	\includegraphics[width=\columnwidth]{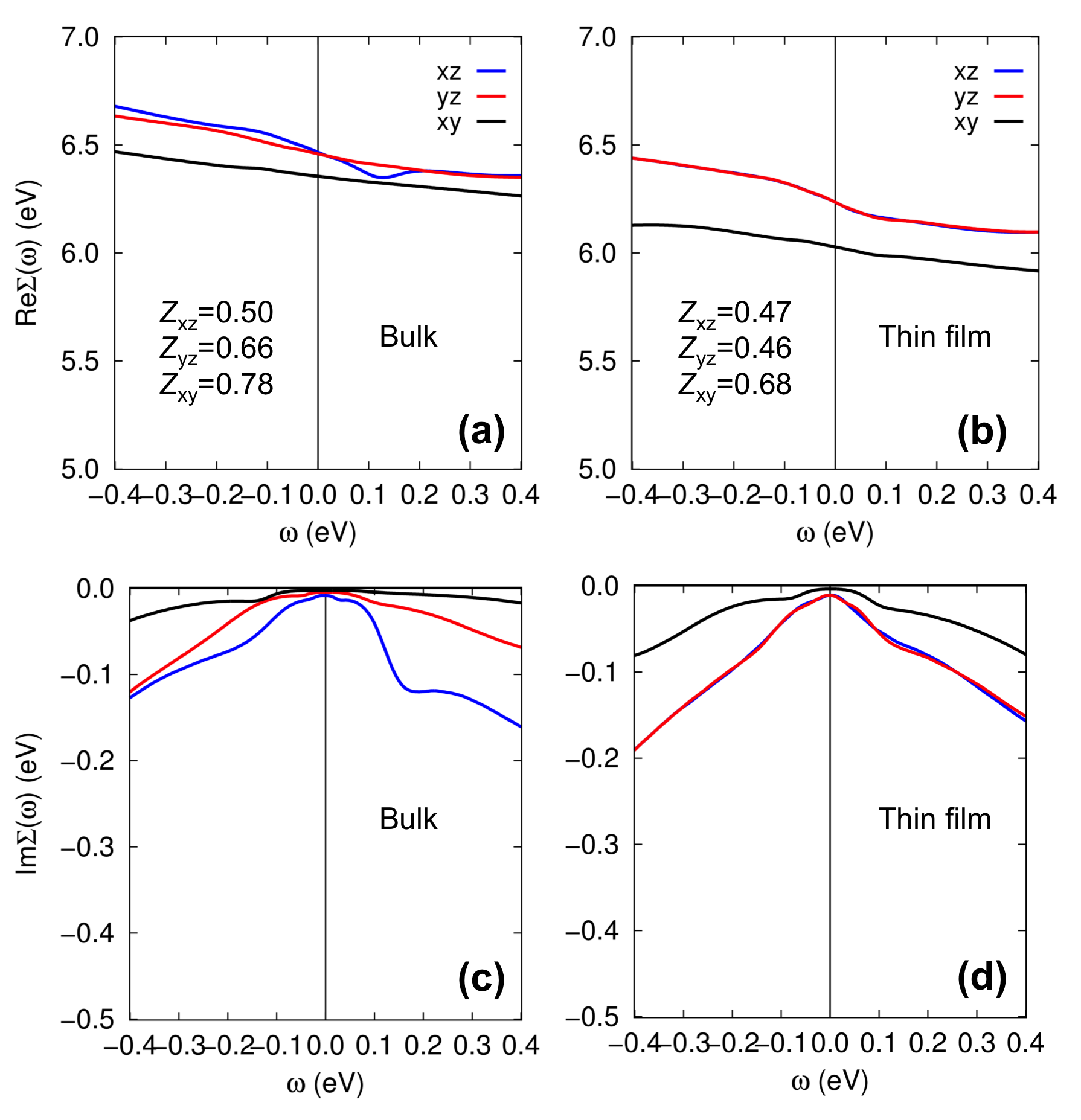}
	\caption{DFT+DMFT \mbox{self-energy} for the bulk and the thin film at $T$=58 K. (a) Real part, bulk; (b) Real part, thin film; (c) Imaginary part, bulk;  (d) Imaginary part, thin film. Corresponding renormalization factors $Z$ are listed in the figures.
		\label{fig:Self_energies}
	}
\end{figure}

The total DOS obtained by DFT+DMFT is shown in Fig.\ref{fig:Density_of_states}(a,b). One can notice that DFT+DMFT calculations predict metallic behavior for both the bulk and the thin film. Only moderate effects of electronic correlation with renormalization of $Z\approx 0.5$ are observed ($Z=(1-\frac{\partial Re\Sigma(\omega)}{\partial \omega})^{-1}|_{\omega \rightarrow 0}$), as obtained by the \mbox{self-energy} results shown in
Fig.\ref{fig:Self_energies}. The thin film is slightly more correlated with smaller values of $Z$, but no major difference between the thin film and the bulk is observed, in contrast to the experimental results.

According to the orbitally resolved DFT+DMFT DOS in Fig.\ref{fig:Density_of_states}(c,d), due to the larger bandwidth of the $xy$ orbital than that of the $xz/yz$ orbitals, the occupancy of the $xy$ orbital (1.50) is smaller than that of the $xz/yz$ orbitals (1.58) in the thin film, which is consistent with the Rh $L_2$ edge XLD results. As depicted in Fig.\ref{fig:Density_of_states}(e-g), XLD is defined as the difference of XAS measured by using incident \mbox{x-rays} with $E//a$ and $E//c$, where $a$ and $c$ are the in-plane [100] and the out-of-plane [001] directions, respectively. Since the XAS intensity at the Rh $L$ edge is proportional to the number of 4$d$ holes, a positive XLD signal indicates a preferred occupation of out-of-plane 4$d$ orbitals and less occupation of in-plane 4$d$ orbitals. Note that the sign change in the $L_3$ edge XLD spectrum could often be observed in other systems as well, while the spectrum at the $L_2$ edge can usually reflect the orbital occupation more unambiguously~\cite{38_pesquera2012surface}. These experimental and calculation results are consistent with the biaxial compressive strain from the SrTiO$_3$ substrate~\cite{16_nichols2017electronic}.

\begin{figure}[t]
	\includegraphics[width=\columnwidth]{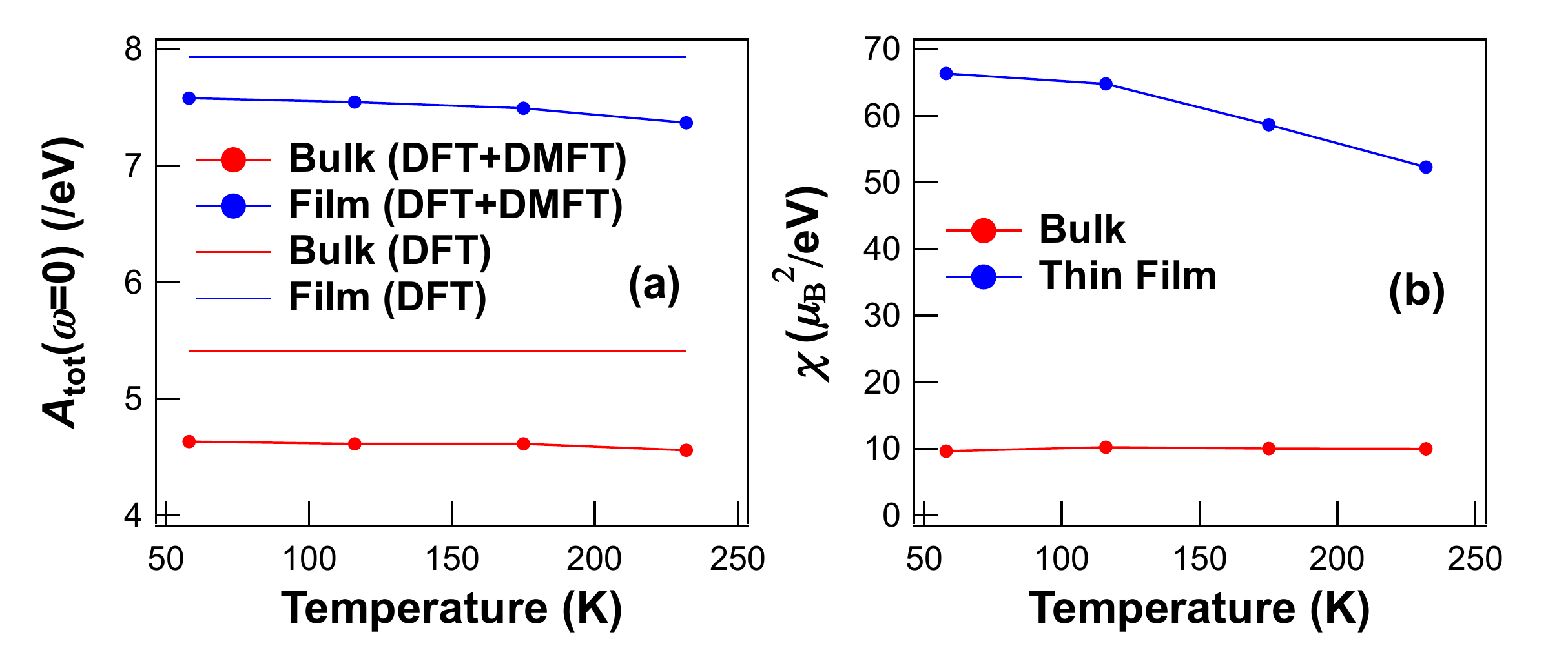}
	\caption{(a) Temperature dependent total DFT+DMFT DOS
		at $E_F$ for the bulk and the thin film. The total DOS calculated by DFT for the bulk and the thin film are also plotted for comparison. (b) Temperature dependent magnetic susceptibility
		calculated by DFT+DMFT for the bulk and the thin film. 	
		\label{fig:Susceptibilities}
	}
\end{figure}

On the other hand, in contrast to the strong orbital anisotropy of spectral weight in between the $xy$ and the $xz/yz$ orbitals, the orbital dependence of quasiparticle renormalization is not as strong as that in Sr$_{2}$RuO$_{4}$~\cite{31_mravlje2011coherence,39_tamai2018high}, which is consistent with the claim in Ref.~\cite{40_yamaura2004ferromagnetic} that correlation effects are weaker in SrRhO$_{3}$ than in perovskite ruthenates.
The electronic correlation changes the effective energy level of the $xz$, $yz$, and $xy$ orbitals.
In the DFT calculation, the center energy of the $xy$ orbital is 618~meV higher than that of the $xz/yz$ orbitals in the SrRhO$_3$ thin film.
In the DFT+DMFT \mbox{self-energy} of the thin film (Fig.\ref{fig:Self_energies}(b)), the $xz/yz$ orbitals are shifted up with respect to the $xy$ level for 212~meV,
as shown by the difference of the real part of \mbox{self-energy} at zero energy.
As a result, the effective energy level of $xy$ with respect to that of $xz/yz$ is reduced from 618~meV to 404~meV
by electronic correlation in DFT+DMFT calculations.
Since the Van-hove peaks of the $xz/yz$ orbitals in DFT are close to $E_F$ within 30~meV (Fig.\ref{fig:Density_of_states}(a,b)), the correlation-induced shift-up of the $xz/yz$ orbitals gives rise to the reduction of the DOS at $E_F$ in DFT+DMFT, as shown in Fig.\ref{fig:Susceptibilities}(a).
Both the bulk and the thin film show a similar trend that the total DOS at $E_F$ is reduced by correlation, which is qualitatively consistent with the small electronic component of the experimental specific heat of SrRhO$_3$~\cite{13_yamaura2001enhanced}.
It is noteworthy that the $xz/yz$ orbitals have a larger DOS at $E_F$ than the $xy$ orbital in the thin film. Meanwhile, the result that $Z\approx 0.5$ for all orbitals implies that electronic correlation
is not so sensitive to the value of the DOS at small energies, which is different from Hund’s metals such as ruthenates and iron-based superconductors~\cite{31_mravlje2011coherence,41_georges2013strong}.

%%%%%%%%%%%%%%%%%%%%%%%%%%%%%%%%%%%%%%%%%%%%%%%%%%%%%%%%%%%%%%%%
%%%%%%%%%%%%%%%%%%%%%%%%%%%%%%%%%%%%%%%%%%%%%%%%%%%%%%%%%%%%%%%%

We also calculated the uniform magnetic susceptibility of both the bulk and the thin film, as shown in Fig.\ref{fig:Susceptibilities}(b). In contrast
to the \mbox{self-energy}, the calculated magnetic susceptibility does show a substantially different behavior in the thin film. The magnetic susceptibility of the thin film is 6 to 7 times larger and exhibits a stronger temperature dependence than that of the bulk case, in contrast to the nearly temperature independent total DOS in the bulk and the thin film (Fig.\ref{fig:Susceptibilities}(a)).
The difference in the calculated magnetic susceptibility for the bulk and the thin film can be understood as follows. First, the larger total DOS at $E_F$ in the thin film with respect to that in the bulk (Fig.\ref{fig:Density_of_states}(b)) gives rise to a larger magnetic susceptibility.
Second, the stronger electronic correlation of the $xz/yz$ orbitals in the thin film compared to that in the bulk (Fig.\ref{fig:Self_energies}(a-d)) gives rise to a
larger magnetic instability in the thin film. Third, the sharper slope in the DOS of the $xz/yz$ orbitals in the thin film compared to that in the bulk (Fig.\ref{fig:Density_of_states}(c,d)) gives rise to a
stronger temperature dependence of the magnetic susceptibility in the thin film.
These results are inherited from the tetragonal symmetry of the lattice of the thin film, which gives rise to the presence of the Van-hove singularity
of the $xz/yz$ orbitals near Fermi level. 
%%%%%%%%%%%%%%%%%%%%%%%%%%%%%%%%%%%%%%%%%%%%%%%%%%%%%%%%%%%%%%%%
%%%%%%%%%%%%%%%%%%%%%%%%%%%%%%%%%%%%%%%%%%%%%%%%%%%%%%%%%%%%%%%%

In our photoemission results, it is clearly shown that there is a negligible
temperature dependence of the DOS between 80~K and 300~K. The present
DFT+DMFT results with similar renormalization for both the bulk and
thin film suggest that if there is a real transition of the electronic structure, it will not be a simple metal-insulator transition with a
Mott gap. The larger magnetic susceptibility in the thin film compared to that in the
bulk implies that the SrRhO$_3$ thin film has a much stronger intrinsic
instability towards magnetically ordered phases. This magnetic instability is mainly induced by the
anisotropy of the crystal environment, such as crystal field symmetry and
bandwidth anisotropy. However, whether this
larger magnetic susceptibility is a side effect (or indicator) of some
actual electronic instability that in turn is responsible for the
experimentally observed neglibible DOS at $E_F$ is an open
question.  

Earlier DFT calculations reported the occurrence of
an AFM state in SrRhO$_3$ thin films~\cite{16_nichols2017electronic}. We
investigated the possibility of magnetic ordering by DFT+$U$
calculation and found that we need $U>5$~eV for the stabilization of
the C-type AFM state, which is too large for the 4$d$ shell~\cite{42_deng2016transport,43_fang2004orbital}.
We also conducted RXMD
experiments at the Rh $L$ edges to attest the existence of AFM ordering
peaks. Due to the restricted $Q$ range that the Rh $L$ edge \mbox{x-ray} (around
3 keV) can reach, $Q$ vectors of (0 0 0.5) (A-type), (0.5 0.5 1)
(C-type) and (0.5 0.5 0.5) (G-type) were investigated at 30~K but no
observable diffraction appeared within the detection limit.

\section{About Negligible DOS at $E_F$ in \NoCaseChange{SrRhO$_3$} Thin Film: SOC, Ordering and Beyond}

%%%%%%%%%%%%%%%%%%%%%%%%%%%%%%%%%%%%%%%%%%%%%%%%%%%%%%%%%%%%%%%%
\begin{figure}[t]
	\includegraphics[width=\columnwidth]{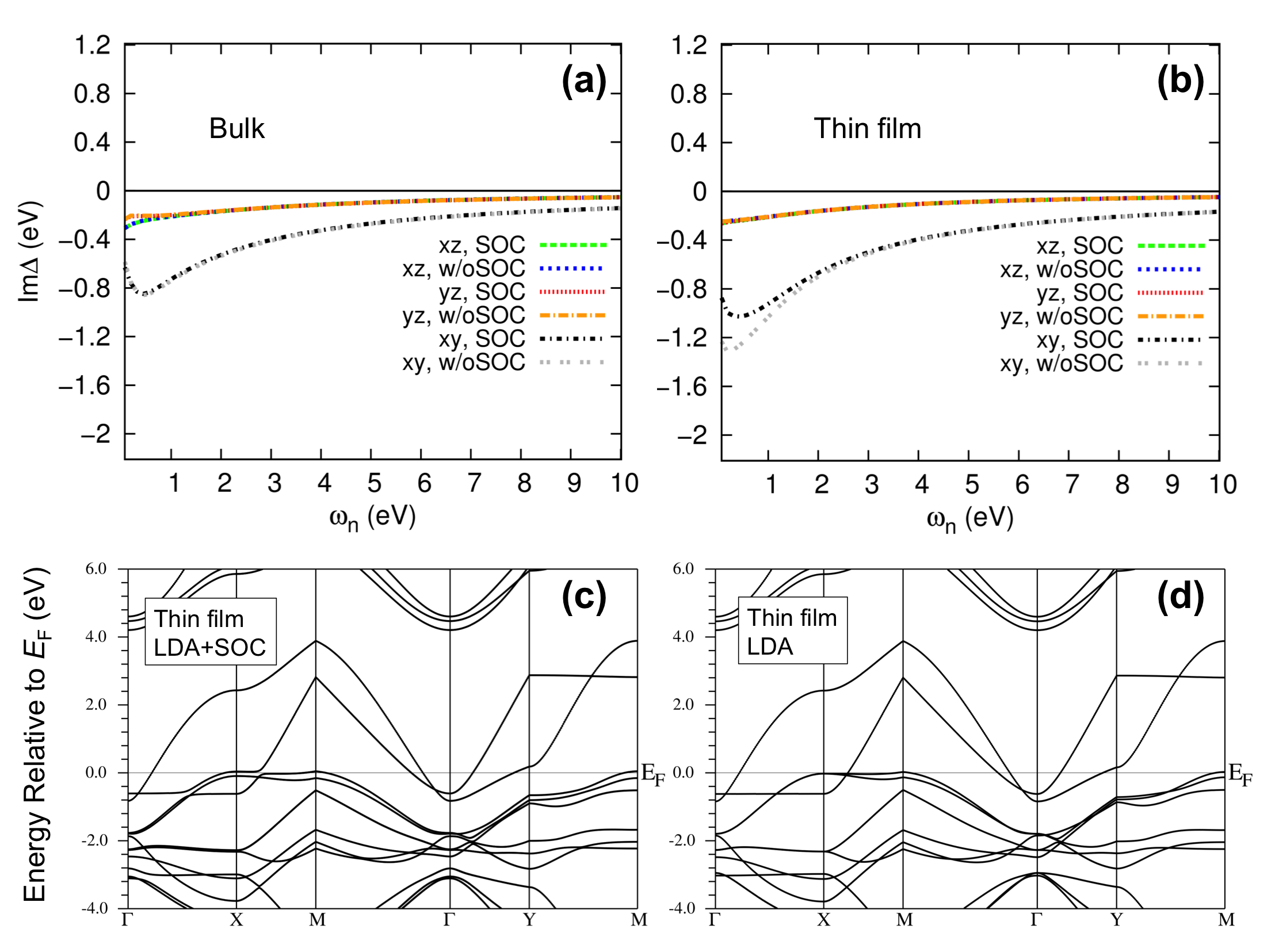}
	\caption{Hybridization functions in the initial step of DFT+DMFT calculation for (a) the bulk and (b) the thin film. $\omega$\textsubscript{n} is the Matsubara frequency. Band structures for the thin film calculated by DFT (c) with and (d) without SOC.
		\label{fig:Spin_orbit_coupling}
	}
\end{figure}
%%%%%%%%%%%%%%%%%%%%%%%%%%%%%%%%%%%%%%%%%%%%%%%%%%%%%%%%%%%%%%%%
In principle, a possible origin of the small value of measured DOS at $E_F$ and the
absence of a coherent peak could be SOC, which can play an important
role in $t_{2g}^5$ iridates~\cite{44_kim2008novel,45_moon2008dimensionality}. If SOC were
strong enough to split the $J$\textsubscript{eff}$=1/2$ and
$J$\textsubscript{eff}$=3/2$ states significantly, one could expect
that an insulating behavior would be promoted by the half-filled
$J$\textsubscript{eff}$=1/2$ band. We are unable to run the DFT+DMFT
calculation in the presence of SOC because of the fermionic sign
problem in the quantum impurity solver~\cite{29_seth2016triqs,30_gull2011continuous}. But to get a first
impression of the possible role of SOC, we calculated the
hybridization functions, which determine the behavior of the DMFT
calculation in the presence of SOC. The results shown in
Fig.\ref{fig:Spin_orbit_coupling}(a,b) imply that SOC moderately
affects the electronic structure of SrRhO$_3$. As also shown by DFT
results with and without SOC in
Fig.\ref{fig:Spin_orbit_coupling}(c,d), in the thin film, due to the
SOC-induced band splitting around $E_F$ (along $\Gamma$ to X line),
SOC reduces the hybridization function of the $xy$ orbital for a small
energy scale of 1 to 2~eV
(Fig.\ref{fig:Spin_orbit_coupling}(b)). Provided the fact that
quasiparticle residue $Z\approx 0.5$ and a small orbital dependence of
the Fermi velocity, we suggest that SOC can not trigger
the metal-insulator transition but might give rise
to a strong magnetic instability in the SrRhO$_3$ thin film. For bulk
SrRhO$_3$, the effect of SOC is even smaller due to the lower lattice
symmetry, as shown in Fig.\ref{fig:Spin_orbit_coupling}(a). Note
that even in SrIrO$_3$, the 5$d$ counterpart of SrRhO$_3$, SOC is
still not strong enough to trigger an insulating behavior~\cite{45_moon2008dimensionality}. Moreover, we can get similar conclusion by analyzing the branching ratio (BR) of XAS results shown in Fig.\ref{fig:Density_of_states}(e,f). The BR between the white-line intensities of Rh $L_3$ and $L_2$ edges is related to the ground-state expectation value of the angular part of SOC~\cite{46_van1988local}. A large deviation from the statistical BR=2 indicates the presence of strong SOC effects. The experimental BR at the Rh $L_{3,2}$ edges is close to the statistical value of 2 (estimated as around 2.3 from Fig.\ref{fig:Density_of_states}(e,f)), indicating weak SOC effects in the SrRhO$_3$ thin film. This is in contrast to the Ir 5$d$ cases where large deviations (BR$>4$) from the statistical value, thus large SOC, have been observed~\cite{47_laguna2010orbital, 48_clancy2012spin, 49_kim2018controlling}.

There is also the possibility of more complicated magnetic or charge
ordering, such as helical magnetic ordering or spin/charge density
waves, which could be responsible for the absence of coherent peak in the
SrRhO$_3$ thin film. Another possible mechanism could be formation of
polarons induced by electron-phonon interaction. These possibilities
should be considered and investigated in future to further clarify the electronic structure of SrRhO$_3$ thin films.

\section{Conclusions}
In summary, we experimentally and theoretically investigated the
effects of electronic correlation in SrRhO$_3$. The photoemission results indicate a
negligible DOS at $E_F$ in the SrRhO$_3$ thin film with little temperature
dependence. We considered SrRhO$_3$ within band-structure calculation
taking into account the electronic correlation with a DFT+DMFT
approach. In our calculation the small DOS at $E_F$ could not be reproduced,
rather a moderately correlated metallic behavior was observed. Our attempts to detect the AFM magnetic diffraction by
experiment and to stabilize magnetically ordered states in the calculations
both failed. But the calculation did reveal an interesting behavior in the
magnetic susceptibility that is substantially larger for the thin
film. This result is mainly induced by the difference of crystal
anisotropy between the bulk and the thin film.
 Further investigations are welcome to identify the
possible magnetic ordering and clarify the effects of SOC in SrRhO$_3$. Moreover, 5$d$
transition metal oxides, such as SrIrO$_3$ or Sr$_2$IrO$_4$, have
strong SOC effects on the electronic correlation~\cite{50_zhang2013effective}. We are also looking forward to
comparative theoretical studies on them.

\section{Acknowledgements}
This work was supported by Grant-in-Aid for JSPS fellows (No. 17F17327). The HAXPES experiments at SPring-8 were performed under the approval of the Japan Synchrotron Radiation Research Institute (Proposal No.~2018B1449). This research used resources of the Advanced Photon Source, a U.S. Department of Energy (DOE) Office of Science User Facility operated for the DOE Office of Science by Argonne National Laboratory under Contract No.~DE-AC02-06CH11357. The work at ORNL was supported by the U.S. Department of Energy, Office of Science, Basic Energy Sciences, Materials Sciences and Engineering Division. M. K. acknowledges support from Grant No.~NSF~DMR-1733071, and grateful to CPHT computer support team. We are thankful to the support and advices provided by A.~Georges, V.~R.~Cooper, S.~F.~Yuk, and A.~Rastogi. And we also acknowledge the support provided by K.~Ikeda, S.~Sakuragi and H.~Kinoshita, as well as enlightening discussion about this work with J.~W.~Kim and H.~Zhou during our beamtime in Advanced Photon Source.

\bibliographystyle{unsrt}
%\bibliography{refs_SrRhO3.bib}

%
\end{document}